\documentclass[12pt]{article}

\usepackage{amssymb}
\usepackage{amsmath}
\usepackage{epsfig}

\usepackage{epsf}
\usepackage{graphicx}
\usepackage{amsfonts}
\usepackage{amsthm}
\usepackage{mathrsfs}

\newcommand{\unit}[1]{\ensuremath{\, \mathrm{#1}}}

\makeatletter
\renewcommand\section{\@startsection {section}{1}{\z@}%
                                 {-3.5ex \@plus -1ex \@minus -.2ex}
                                   {2.3ex \@plus.2ex}%
                                   {\normalfont\large\bfseries}}
\renewcommand\subsection{\@startsection{subsection}{2}{\z@}%
                                   {-3.25ex\@plus -1ex \@minus -.2ex}%
                                     {1.5ex \@plus .2ex}%
                                     {\normalfont\bfseries}}
\renewcommand\subsubsection{\@startsection{subsubsection}{3}{\z@}%
                                   {-3.25ex\@plus -1ex \@minus -.2ex}%
                                     {1.5ex \@plus .2ex}%
                                     {\normalfont\itshape}}
\makeatother

\newcommand{\be}{\begin{equation}}
\newcommand{\ee}{\end{equation}}
\newcommand{\bea}{\begin{eqnarray}}
\newcommand{\eea}{\end{eqnarray}}
\newcommand{\barr}{\begin{array}}
\newcommand{\earr}{\end{array}}

\def\beq{\begin{equation}}
\def\eeq{\end{equation}}
\def\be{\begin{equation}}
\def\ee{\end{equation}}
\def\bea{\begin{eqnarray}}
\def\eea{\end{eqnarray}}

\DeclareRobustCommand{\SkipTocEntry}[4]{}

\begin{document}

\begin{titlepage}

\setcounter{page}{1} \baselineskip=15.5pt \thispagestyle{empty}

\begin{flushright}
SU-ITP-10/13
\end{flushright}
\vfil

\begin{center}
{\LARGE 
Gravity waves and the LHC: \\[1.ex] 
Towards high-scale inflation with low-energy SUSY}

\end{center}
\bigskip\

\begin{center}
{\large Temple He$^1$, Shamit Kachru$^{2,3,\star}$, and Alexander Westphal$^1$}
\end{center}

\begin{center}
\textit{$^1$Department of Physics, Stanford University, Stanford CA 94305}\\
\textit{$^2$Kavli Institute for Theoretical Physics, Santa Barbara CA 93106}\\
\textit{$^3$Department of Physics, University of California, Santa Barbara CA 93106}
\end{center} \vfill

\noindent  
It has been argued that rather generic features of string-inspired inflationary theories with low-energy supersymmetry (SUSY) make it difficult to achieve inflation with a Hubble scale
$H > m_{3/2}$, where $m_{3/2}$ is the gravitino mass in the SUSY-breaking vacuum state. 
We present a class of string-inspired supergravity realizations of chaotic inflation where a simple, dynamical mechanism
yields hierarchically small scales of post-inflationary supersymmetry breaking.  Within these toy models we can easily achieve small ratios between $m_{3/2}$ and the Hubble scale of inflation.  
This is possible because the expectation value of the superpotential $\langle W \rangle$ relaxes from large to small values during the course of inflation.
However, our toy 
models do not provide a reasonable fit to cosmological data if one sets the SUSY-breaking scale to
$m_{3/2} \leq {\rm TeV}$.
Our work is a small step towards relieving the apparent
tension between high-scale inflation and low-scale supersymmetry breaking in string compactifications.

\vfill
\begin{flushleft}
March 22, 2010
\end{flushleft}
\vfill
\rule[-0.1cm]{6.5in}{0.02cm}\\ 
$^\star$ On leave of absence from Stanford University and SLAC.

\end{titlepage}

\newpage
\tableofcontents
\newpage

\section{Introduction}\label{intro}
Inflation \cite{classics} can solve both the horizon and flatness problems of cosmology in an elegant and minimal way (for a recent pedagogical review, see \cite{Baumann09}).   Inflationary theories can also naturally explain the primordial density fluctuations that eventually collapse to give rise to the
large-scale structure we see today.

However, all known classes of inflationary models are potentially sensitive to Planck-suppressed
corrections to the inflaton Lagrangian, which can yield slow-roll parameters of ${\cal O}(1)$, stopping inflation.  While this
sensitivity to high-scale physics is true of all inflationary theories, a particularly stark example of UV sensitivity arises in
so-called ``large-field models," where the inflaton enjoys a super-Planckian excursion in field space during
inflation.  Such models are of special interest because it is only in such models that one may obtain gravitational wave
signatures that are observable in the forseeable future \cite{Lyth} (for a thorough discussion, see \cite{CMBPol}).  But gaining control of the inflaton Lagrangian over this large
range of field space clearly demands detailed knowledge of the structure of an infinite series of potential Planck-suppressed
terms.

One possible UV completion of particle physics and gravity is string theory.  In recent years, as our understanding of the
details of string compactification has grown, it has become a realistic possibility to enumerate precise corrections to
candidate inflaton Lagrangians in various scenarios.   Recent results in this direction include those of \cite{Liam}, where 
possible quantum
gravity corrections to
D3-brane inflation models in warped throat geometries are determined, and those of \cite{monodromy}, where a shift symmetry protects large-field
inflation in theories with high-scale supersymmetry breaking.\footnote{See \cite{axions} for other papers
attempting to use shift symmetries to justify large-field inflation in string theory, and \cite{reviews} for more general reviews.}
The UV sensitivity of all inflation models, and the
especially stark sensitivity of models which predict observable gravitational waves,
provides the principal motivation for trying to embed inflation in string theory. 

A second major paradigm in theoretical physics is supersymmetry (see e.g. \cite{susyreview}).  Many theorists
believe that supersymmetry is the leading candidate to stabilize the Higgs mass and explain the physics of electroweak symmetry
breaking.   String compactifications very naturally give rise to models with low-energy supersymmetry (though this is by no means
known to be a prediction of the framework); the low-energy theory is then a 4D ${\cal N}=1$ supergravity.
Therefore, in evaluating statements about the space of inflationary models in string theory, and in particular
statements that correlate inflationary observables with particle physics observables directly tied to supersymmetry
breaking, it is useful to first consider what
is and is not possible in the context of string-inspired low-energy supergravities.

Recently, a striking claim about the relation between the two most basic observables in inflation and in theories of supersymmetry
and its breaking
has been put forward \cite{Kallosh04,CMB}\footnote{See also~\cite{quibble} which noted tensions cropping up between low-scale SUSY breaking and high-scale inflation.}.  The most fundamental observable in inflation is the scale of inflation 
\begin{equation}
V = 3 M_{\rm P}^2 H^2
\end{equation}
 (or equivalently,
the Hubble constant during inflation, $H$).  
It directly controls the tensor amplitude~\cite{Lyth}, and is a major factor in setting the scale of density perturbations.
The primary observable in any realistic supergravity model is the scale
of supersymmetry breaking, which is captured by the gravitino mass $m_{3/2}$.  Quantitatively,
one has:
\begin{equation}\label{gravitinomass}
	m_{3/2}^2 \approx {e^{K}|W|^2 \over M_{\rm P}^4} \end{equation}	
where $W$ is the expectation value of the superpotential and $K$ is that of the K\"ahler potential (and the two appear above
in a combination which is invariant under K\"ahler transformations, as expected).

The authors of \cite{Kallosh04} study possibilities for inflation in one of the simplest toy models of moduli stabilization and
supersymmetry breaking known in string theory \cite{KKLT}.  They claim that within this class of models, very simple
arguments (which we shall review in section~\ref{KL}) lead to the conclusion that one must have
\begin{equation}
\label{quack}
	H \leq m_{3/2}~.
\end{equation}
The basic extra microscopic requirement that leads to this constraint is that of volume modulus stabilization (as we shall explain in detail in section~\ref{KL}).
This is a ${\it new}$ microscopic requirement that must be considered in inflationary models that arise in an extra-dimensional setting, like that of
string theory; it is ${\it a ~priori}$ only indirectly related to traditional questions of 4D inflaton dynamics, like the flatness of a candidate
inflaton potential.

We note that typical supersymmetric models of particle physics have $m_{3/2} \leq {\rm TeV}$ (sometimes far lower,  coming all the
way down to
$10^{-2}\ {\rm eV}$
in models of low-scale gauge mediation).  In contrast, typical models of inflation have a characteristic energy scale $V$
during inflation that often approaches the GUT scale.  All models with observable gravitational waves predict 
$H \geq 10^{14}\ {\rm GeV}$, and very few models of any sort have been proposed with $H$ smaller than the
values of $m_{3/2}$ typical in low-scale gauge mediation (recall
that one must do baryogenesis etc. sometime ${\it after}$ inflation).  Therefore, the constraint eq.~(\ref{quack}) is 
rather unwelcome.
It has been further argued that while one can (clearly) find more general low-energy Lagrangians generalizing that of \cite{KKLT}, 
that allow one to circumvent eq.~(\ref{quack}), rather significant fine-tuning in the moduli-stabilizing sector is required to obtain models that robustly allow $H \gg m_{3/2}$.


In this paper, we examine the conclusion of~\cite{Kallosh04} by studying possibilities for large-field inflation in low-energy theories that
incorporate the same model of moduli stabilization, but vary the nature of the inflationary
sector.  We do not work in the full framework of string theory, but we do incorporate all of the features of low-energy
string models
that led to the tension in \cite{Kallosh04}.
We find that a wide class of large-field models can nevertheless arise in this framework, with $m_{3/2} \ll H$, and without significant fine-tuning of
the moduli-stabilizing sector.\footnote{For other work focused on related issues, see e.g. \cite{Postma,BaOlkillKL,Covi,KLkillKL,Burgess,Shiu}, and for new
ideas about using the universal supergravity Goldstino multiplet for inflation, see \cite{Luis}.}

We emphasize that the problem described in \cite{Kallosh04}, and solved there only by significant fine-tuning in the moduli-stabilizing sector,
is ${\it different}$ from the problem of obtaining a stringy inflation sector with a flat inflaton potential; it comes instead
largely from a constraint to avoid decompactification of the extra dimensions of string theory.  It is this new problem 
that is the focus of our investigation.

\section{The Kallosh-Linde Problem}\label{KL}

The Kallosh-Linde (KL) problem was originally described as follows (for a more complete discussion, see the original paper \cite{Kallosh04}).\footnote{To simplify the equations presented, we will henceforth set the reduced Planck mass $M_{\rm P} = 1$.} 
For concreteness, we imagine working in type IIB string theory on a Calabi-Yau orientifold, and
denote its volume modulus field by $T$, with $\sigma \equiv \text{Re}\,T$ (and the imaginary part being comprised of an axion). In the scenario
of \cite{KKLT}, the K\"ahler potential $K$ and the superpotential $W$ (in the effective theory below the scale where complex structure
moduli are stabilized by fluxes) take the form
\begin{eqnarray}
K&=&-3\ln(T+\bar T)\\ && \nonumber\\
W &=& W_0 + Ae^{-aT}~.\label{Wis}
\end{eqnarray}
$W_0$ is the value of the flux superpotential at the minimum for complex structure moduli, and the exponential term in $W$ arises from non-perturbative effects.
The resulting scalar potential has an AdS minimum, which is supersymmetric. The $F$-terms vanish, and denoting
the value of the superpotential in the supersymmetric AdS vacuum by $\langle W\rangle_0$ (see eq.~(14) of \cite{KKLT}), the potential has a depth of
\begin{equation}\label{VAdS}
	|V_{AdS}| = 3e^K|\langle W\rangle_0|^2.
\end{equation}

One then further incorporates some effects of supersymmetry breaking to lift the AdS minimum to a metastable de Sitter minimum.  There are many ways that one
can imagine incorporating supersymmetry breaking in these constructions; for a discussion of some of these, see the reviews \cite{DKReview}.   The upshot in many
cases is that one obtains a correction to the potential of the form
${\Delta V} \sim {C\over \sigma^2}$, where $C$ can be parametrically small in string or Planck units.  This additive form of the correction is obviously a crude model of a more intricate interaction between the SUSY-breaking sector
and the other dynamics.  Such a form should be (approximately) justified in cases where the SUSY breaking sector only couples energetically to other fields by parametrically smaller 
(e.g. Planck-suppressed) terms.

 For appropriate choices of $C$, this can ``uplift" the AdS minimum at $\sigma_{min}$ to a de Sitter minimum (a similar but different power-law dependence of $\Delta V$ on the volume modulus will also work; typical sources of energy density in string theory indeed scale in this way with the
volume modulus, when one works in 4D Einstein frame).   The correction factor is small enough that the new minimum occurs at $\sigma_0 \approx \sigma_{min}$. The smallness of the correction also guarantees that the barrier height preventing
decay of the de Sitter vacuum to the vacuum with $V=0$ at $\sigma=\infty$, denoted $V_B$, is 
\begin{equation}
\label{barrier}
	V_B \simeq |V_{AdS}|.
\end{equation}

Given the value of the present-day cosmological constant, the potential at the end of inflation must effectively vanish (giving rise to a vacuum energy density of order $10^{-120}$ in Planck units) at the minimum.  The gravitino mass is given by eq.~(\ref{gravitinomass}). 
In the simplest case where SUSY is broken by the $F$-term of some chiral multiplet $Z$,
\begin{equation}
	V = G^{Z\bar Z}|F_Z\vert^2 - 3e^K |W|^2,
\end{equation}
so a vanishing $V$ implies that $|F_Z|^2\equiv |e^{K/2}D_Z W|^2 = 3e^K|\langle W\rangle_0|^2$ (here we assume that $Z$ is a canonical field so $G^{Z\bar Z} = 1$). Hence the $F$-terms, which measure the scale of SUSY breaking, are of the same order as $e^{K/2}|\langle W\rangle_0|= m_{3/2}$, and so the gravitino mass is a direct measure of the scale of SUSY breaking.  This remains true in more complicated models. To summarize, the scale of SUSY breaking is directly tied to $m_{3/2}$ and
\begin{equation}\label{gravitinomass2}
	m_{3/2}^2 = e^K|\langle W\rangle_0|^2 
\end{equation}
when the cosmological constant is very small (as it is today).

On the other hand, we now argue that $V_B$ also imposes an upper bound on the magnitude of $H^2$.
Let us modify the scenario of \cite{KKLT} to include inflation, by adding an inflaton field $\Phi$.
We assume that $T$ modulus stabilization works in the same way as above at the end of inflation, when $\Phi$ vanishes.
It follows from this that the final $V_B$ is still as in eq.~(\ref{barrier}). 

Now, let us consider the effects of the inflaton contributions to the potential during inflation, when
\begin{equation}
\label{potential}
	V = e^K(G^{\Phi\bar\Phi}|D_\Phi W|^2 +G^{T\bar T} |D_T W|^2 - 3|W|^2) + \frac{C}{\sigma^2}.
\end{equation}
We note that if the new terms due to the inflaton in eq.~(\ref{potential}) are much larger than the barrier height $V_B$, we
can expect a problem with decompactification, since for all known inflaton candidates $V(\Phi) \sim e^K G^{\Phi\bar\Phi}|D_{\Phi}W|^2$
vanishes as a power of $1/\sigma$ at large $\sigma$.  (A typical value of the power is $1/\sigma^3$ from the prefactor $e^K$).
In other words, the $e^K |D_\Phi W|^2$ term is effectively an uplifting term, similar in functional form to the $\overline{D3}$ contribution in the scenario
of \cite{KKLT}. Empirically, it has been argued that to prevent the $|F_\Phi|^2$ terms from overuplifting the potential and destroying the minimum in the volume modulus field, we need \cite{Kallosh04}
\begin{equation}\label{Ftermbound}
	e^K|D_\Phi W|^2 \lesssim \mathcal O(10)V_B.
\end{equation}
Thus, as $H^2 = \frac{V}{3}$ during inflation,
\begin{equation}
	H^2 = \frac{V}{3} \sim e^K|D_\Phi W|^2 \lesssim \mathcal O(10)V_B \simeq \mathcal O(10)|V_{AdS}|, 
\end{equation}
where the last approximate equality is from eq.~(\ref{barrier}).  So, under this set of assumptions, $V_B$ is related to both the
gravitino mass, and the maximal possible scale of inflation.

It is now easy to formulate the KL problem. If we assume that the $\sigma$ field remains at its minimum during inflation, then the scale of inflation is given by (using eq.~(\ref{VAdS}))
\begin{align}
	H^2 \lesssim {\cal O}(10)V_B \simeq \mathcal O(10)|V_{AdS}| \sim e^K|\langle W\rangle_0|^2 \sim m_{3/2}^2,
\end{align}
which is just eq.~(\ref{quack}) from the introduction (with $M_{\rm P} = 1$).   This equation leads to the statement in \cite{Kallosh04} that, due to the need to maintain
stability of the volume modulus during inflation, inflationary models in string theory should generically be expected to satisfy
\begin{equation}
	H_{\text{inflation}} \leq  m_{3/2}^{\text{today}}.
\end{equation}
This ties the scale of SUSY breaking to the scale of inflation. For many high-scale inflationary models this yields $m_{3/2} \sim 10^{10} \unit{GeV}$ in the simplest scenario of \cite{KKLT}, many orders of magnitude greater than the 1~TeV value predicted by typical supersymmetric models.   The KL problem 
suggests that it may be difficult to find inflation models that can accommodate both a potential future observation of tensor modes from inflation and a light gravitino. 

Kallosh and Linde did propose a way to circumvent this problem. They noted that eq.~(\ref{quack}) was derived from the fact that the post-inflationary near-Minkowski de Sitter minimum cannot be further uplifted by too large a factor (more specifically, the uplifting cannot greatly exceed $V_B$) \cite{Kallosh04}. Hence, if it is possible to free the uplifting limit from $V_B$, then eq.~(\ref{quack}) becomes invalid, eliminating the problem. To accomplish this, Kallosh and Linde proposed to add a second exponential in the $\sigma$ field to $W$, thus using the racetrack mechanism to stabilize $\sigma$. By choosing the coefficients of the exponentials carefully, they were able to completely
decouple the potential barrier height from the scale of uplifting. Nevertheless, although this model indeed resolves the problem, it requires significant fine-tuning (for reasons which are ${\it distinct}$ from the typical need to achieve a flat inflaton potential; this
tune is invoked simply to avoid decompactification during inflation).

There is, however, a different approach that can be taken to circumvent the KL problem. Rather than trying to free the uplifting limit from $|\langle W\rangle_0|$, we simply allow $\langle W \rangle=\langle W(\Phi)\rangle$ to vary as a function of the inflaton $\Phi$ during the last 60 e-folds of inflation. During inflation, we imagine that $\langle W \rangle$ is quite large, and the effective barrier to decompactification is high. However,  at the end of inflation, $\langle W \rangle = \langle W\rangle_0$ is also tied to the scale of SUSY breaking through $m_{3/2}$ by eq.~(\ref{gravitinomass2}).   Thus, we are led to search for models where, during the final 60 e-folds, $\langle W \rangle$ naturally decreases by several orders of magnitude.    If we can find such models where $\langle W \rangle$ is large during inflation, but compatible with SUSY breaking at intermediate scale or below at the end of inflation (so $m_{3/2} \leq {\rm TeV}$), then we would have dynamically overcome the KL problem. Our task in the next section is to write down such a toy model.

\section{Large field inflation with small gravitino mass}\label{KLcure}

In this section, we proceed to write down large-field inflation models which are generalizations of chaotic inflation \cite{Andrei}
with $\varphi^{2n}$ potential.  These models are designed
to avoid decompactification even at large vevs of $\varphi$, and the expectation value of the superpotential $\langle W \rangle$ varies by many
orders of magnitude during inflation.  As a result, the final value of $m_{3/2}$ can be much less than the Hubble scale during inflation.
Because the discussion is somewhat detailed, here we provide an overview of our strategy.

We begin in section~\ref{toy}  by writing down the simplest class of models we have found.   They include one additional field $X$ beyond the minimal content
one might expect (the $T$ modulus and the inflaton $\Phi = \eta + i \varphi$) in any discussion of the KL problem.  This additional field $X$ is needed to avoid very general
constraints on large-field inflation in supergravity, discussed in the insightful paper of Kawasaki, Yamaguchi and Yanagida \cite{sugrachaotic}.
The same field allows us to overcome other detailed problems with keeping the $T$ modulus stable during inflation, which would also pose obstacles in a
large-field model
with only $T$ and the inflaton field $\Phi$.  We explain these general constraints in detail in section~\ref{nogo}, using our toy model of section~\ref{toy} as an illustration. 
 In section~\ref{whinnie}, we then scan over the range of parameters
that are accessible in this class of models, exhibiting many models that have $H \gg m_{3/2}$.

\subsection{A toy model}\label{toy}

We begin by writing down the K\"ahler potential and superpotential of our toy model. As before, we take $M_{\rm P} = 1$.
\begin{eqnarray}\label{KWdef}
	K &=& \frac{1}{2}(\Phi + \bar\Phi)^2 + X\bar X - \gamma(X\bar X)^2 - 3\log(T+\bar T)\nonumber\\  &&\nonumber\\
	W &=& W_0 \,g(X) + \alpha \,f(X)\,\Phi^n + e^{-aT}\\ && \nonumber\\
	&& {\rm with:}\;\; g(X)=1+{\cal O}(X)\quad{\rm and}\quad f(X)=b+X+{\cal O}(X^2)\nonumber
\end{eqnarray}
Here, $\Phi=\eta+i\varphi$ is the inflaton, $X$ is a chiral multiplet, and $T$ is the modulus field. The inclusion of the quartic $-\gamma(X\bar X)^2$ term in the K\"ahler potential results in 
\begin{equation}
	K_{X\bar X}^{-1} = (1 - 4\gamma X\bar X)^{-1} \simeq 1 + 4\gamma X\bar X~.
\end{equation}
This effectively produces a mass term for $X$ of order $\sim |F_X|^2$ in the scalar potential, and forces the $X$ field to stay near the origin until inflation has ended at $\Phi\sim 1$. The coefficient of the quartic term in $K$ we take for naturalness to be $\gamma={\cal O}(1)$.   Higher order terms in $X\bar X$ could be added
to $K$ (with the expected ${\cal O}(1)$ coefficients) and would not change our discussion. 
Similarly, one could replace the ${1\over 2} (\Phi + \bar\Phi)^2$ term above with a more general function $F(\Phi + \bar \Phi)$ in the K\"ahler potential; the
only important point is that $F$ should depend only on the ${\it real ~part}$ of $\Phi$.   The possible higher order powers of $(\Phi + \bar \Phi)$ will drop
out in all of our considerations below, because $\eta = {\rm Re}(\Phi)$ is frozen at zero during inflation when $\varphi$ has a large expectation value.

Jumping ahead, we note that during inflation the $F$-terms develop a hierarchy
\beq
F_X\sim e^{K/2}\langle W\rangle\sim e^{K/2}\alpha \Phi^n\quad,\quad F_\Phi\sim\frac{F_X}{\Phi}\quad,\quad F_T\sim\frac{F_X}{T}
\eeq
implying $F_X$-domination during the inflationary phase (which occurs at large values of $\varphi$). Then the dominant term in the scalar potential during inflation is
\beq
V_{inf.}(\varphi)\sim |F_X|^2\sim\alpha^2\varphi^{2n}~.
\eeq
This form of the potential can be protected for large values of $\varphi$ if there is a suitable shift symmetry broken only by $\alpha$; we discuss naturalness issues
below.
The dominant F-term $F_X$ also yields, through the quartic $X$ self-coupling in the K\"ahler potential, a $\varphi$-dependent mass for $X$ given by $m_X^2\sim |F_X|^2\sim V_{inf.}(\varphi)$.  This in turn 
guarantees that $\langle X\rangle \simeq 0$ during inflation, as stated before.  Similarly, $\eta$ is frozen to zero by the large mass it receives from the terms
$\sim e^K \alpha^2 \varphi^{2n}$ in the scalar potential.

Regarding the constants $a$ and $b$, we take $b \in \left[1/4\;,1/\sqrt{2}\,\right]$, and we have assumed the non-perturbative dynamics to arise from gaugino condensation, say on a stack of D7-branes in a warped IIB flux compactification, which gives $a = \frac{2\pi}{N}$. $\alpha$ determines the scale of inflation, and is eventually fixed by 
matching the density perturbations to data if the inflaton itself is chosen to seed the primordial curvature perturbation.

We know that  $|\langle W\rangle_0|$ is of the order $|W_0|$, so if we choose $|W_0|$ to be small,  we will have low-scale SUSY breaking. We should now
choose our constants so that the initial $|W_i|\equiv |\langle W(\varphi_{60},T(\varphi_{60}),X(\varphi_{60})) \rangle|\ll 1$, at the value $\varphi_{60}$ of the inflaton corresponding to 60 e-folds before the end of inflation, is many orders of magnitude larger than $|W_0|$, while the modulus field remains stabilized.  Naturally, not all choices of parameter sets will preserve the (instantaneous) minimum for $T$, and we must derive conditions on the allowed values of $n$ and $W_0$. The constraints on these parameters arise from arguments given in \cite{Kallosh04} (see section~\ref{KL}): The $F$-terms in $V$ act effectively as an uplifting term. In order to prevent decompactification, we must have, by equations~(\ref{VAdS}), (\ref{barrier}), and (\ref{Ftermbound}) with the inclusion of the $F_X$ term,
\begin{equation}\label{norunaway}
	|F_\Phi|^2 + |F_X|^2 \lesssim  \mathcal O(10)3e^K|\langle W \rangle|^2.
\end{equation}
We neglected the $F_T$ term as it is dominated by either $F_\Phi$ or $F_X$. Rewriting the above expression, we have
\begin{equation}
\label{fconstraint}
	\frac{\sqrt{|F_\Phi^2| + |F_X^2|}}{\sqrt{3}e^{K/2}|\langle W \rangle|} \sim \mathcal O(1),
\end{equation}
where we replaced $\mathcal O(10)$ with $\mathcal O(1)$ to be conservative in our estimates. We will see that applying this generic relation to our specific model will produce a constraint on $n$.

\subsubsection*{Naturalness}

Let us now justify the form of the superpotential above; we claim that it can be made natural in the sense of 't Hooft.  We imagine that there is an R-symmetry under which the inflaton field $\Phi$ carries R-charge
${2\over n}$, with $X$ neutral.  $W_0$ itself serves as a spurion of R-symmetry breaking as well.  Therefore, the most generic superpotential consistent
with the R-symmetry can have general functions $f(X)$ and $g(X)$ multiplying $\Phi^n$ and $W_0$.  
In this setup, the $\Phi$ field also possesses a Nambu-Goldstone-like shift symmetry, in that 
\begin{equation}
\Phi \rightarrow \Phi + iC
\end{equation}
with $C$ a real constant, is a symmetry of the K\"ahler potential.  Therefore, for simplicity, we have also pulled an overall small
coefficient $\alpha$ out of the superpotential term $\sim \Phi^n$.  Small $\alpha$ is perfectly natural, since $\alpha$ will serve as the spurion of shift-symmetry breaking.
While there are famously no ${\it exact}$ global symmetries in quantum gravity \cite{Banks,Susskind}, one can sometimes find such shift symmetries which are
protected up to the level of sufficiently small non-perturbative corrections \cite{monodromy}.   That is, the leading symmetry breaking $\sim \alpha$ is generated
dynamically (with the small parameter $\alpha$ arising naturally either through warping, dimensional transmutation, or instanton effects), while any further symmetry breaking
is assumed to be small enough to be negligible for our purposes.

Because of the shift symmetry in $\Phi$ (and our assumption about the nature of the corrections above, which is justified in at least in some stringy large-field models \cite{monodromy}), the inflaton's Lagrangian is ``immune" to corrections even over super-Planckian distances, so we do not need to worry about the slow-roll conditions being destroyed by such corrections. In particular, writing out the components of the inflaton chiral supermultiplet
\begin{equation}
\Phi = \eta + i\varphi~,
\end{equation}
we see that $\varphi$ can be arbitrarily big without affecting $K$. Hence, we can explore large-field inflation, by letting the inflaton be $\varphi$. For now, we are keeping both $n$ and $W_0$ as free parameters.  Our goal is to choose values for them such that there is a hierarchy between $\langle W \rangle$ 
at the start of inflation and $\langle W\rangle_0$ (the superpotential at the end), while avoiding decompactification by satisfying eq.~(\ref{fconstraint}).\footnote{Here, we are only keeping the volume modulus $T$ in our
effective field theory, while integrating out the complex structure moduli, whose stabilization by fluxes is assumed to generate $W_0$.   We must make certain that they
are massive enough to justify integrating them out at the relevant scales.  This is true, because for compactification volumes $R^6\sim {\cal}(10\ldots 100)\alpha'^3$, the mass and energy scales of the flux-induced moduli potential satisfy $M_{\rm P}^2m^2_{mod}\sim U_{mod}\sim\alpha'^2/R^6\sim {\cal O}(10\ldots 100)\times H_{inf}$ even for large-field inflationary models having $H_{inf}\sim 10^{14}\,{\rm GeV}$. This modest hierarchy allows us to infer shifts $\delta \sigma$ of the heavy flux-stabilized moduli $\delta \sigma\sim V_{inf}/U_{mod}$, which implies that corrections to the inflationary slow-roll parameters due to shifts of these moduli vevs scale like $\delta\eta\sim\eta V_{inf}/U_{mod}.$ For a more detailed argument to this effect, see e.g. the discussions in~\cite{monodromy}.}

Since we will actually find that $\langle X \rangle \ll 1$ both during and after inflation in our models, we will only need to keep the first term in the
Taylor expansion of $g(X)$ and the first two terms in the expansion of $f(X)$.  Inclusion of further terms (with generic coefficients) would not change our conclusions.  The reason the first two terms in $f$ are relevant will become clear below and in section~\ref{nogo}.

\subsubsection*{Dynamics of the volume modulus $T$ during inflation}

We pause here to recapitulate the dynamics of ``uplifting" the AdS vacuum for $T$  as in \cite{KKLT}, to see why satisfying the constraint eq.~(\ref{fconstraint}) is sufficient to guarantee the continued existence of an uplifted minimum for $T$. We start by choosing the parameters of the setup such that the eventual minimum for $T$ at large $\varphi$ occurs at $\sigma={\rm Re}\,T\gg 1$. Within that regime, the fact that the main inflationary uplifting comes from $|F_X|^2=e^K|D_XW|^2\sim 1/\sigma^3$ acting like a $\overline{\rm D3}$-brane with respect to the $T$-dynamics,\footnote{Strictly speaking, this is the energetic scaling of an ${\it unwarped}$ anti-D3 brane.  The energy
of a $\overline{\rm D3}$
in a warped throat, which appears in many scenarios, has ${1\over \sigma^2}$ scaling with an exponentially small prefactor, as discussed in section 5.1 of  \cite{KKLMMT}.  That is why we have used this latter form
for the ``uplifting term" throughout our discussion.} shows us empirically that the actual $T$-minimum produced sits at
\beq
T_{min}(\varphi)\simeq T_0\,:\;\left.D_TW(\varphi)\right|_{T_0}=0
\eeq as long as this minimum of $T$ is not very close to disappearing into a barrier-less inflection point. We can use this to estimate the size of the terms in the superpotential at this eventual minimum for $T$. Given that as argued above $D_TW(\varphi)\simeq 0$ at $T=T_{min}(\varphi)$ we have that
\bea
D_TW(\varphi)\simeq 0\quad&\Rightarrow&\quad e^{-a T_{min}(\varphi)}\simeq \frac{1}{1+\frac{2}{3}a T_{min}(\varphi)} W_{0,eff.}(\Phi)\nonumber\\ &&\\
&&\quad {\rm where:}\quad W_{0,eff.}(\Phi)\equiv W_0+\alpha(b+X)\Phi^n\quad.\nonumber
\eea
Thus demanding
\beq |W_{0,eff.}(\Phi)|\equiv |W_0+\alpha(b+X)\Phi^n|\ll1~\forall~|\varphi|=|{\rm Im}\,\Phi|<\varphi_{60}
\eeq
and $a<1$ together are sufficient to ensure that the uplifted $T$-minimum occurs at $\sigma={\rm Re}\,T\gg 1$ and $a\sigma \gg 1$ if it occurs at all.  This guarantees
the validity of both the supergravity approximation and the one-instanton approximation that we use.

Under these conditions, the existence of a minimum for $T$ will be determined by the competition between $|D_TW|^2$, which is trying to relax to zero close to the
location of the old AdS minimum, and $|F_X|^2=e^K|D_XW|^2\sim 1/\sigma^3$, which is adding a positive contribution to the energy that vanishes like a power law at large volume. 
This ultimately results in the condition eq.~(\ref{fconstraint}) for avoiding destruction of the minimum for the $T$-modulus.

Finally, we note that the fact that we can work in the regime of validity of supergravity in the one-instanton approximation, simplifies our further estimates.
In this regime, when estimating the magnitudes of the $F$-terms $F_\Phi$, $F_X$ and of the superpotential $W$, we can neglect the non-perturbative term $\sim e^{-aT}$ in
 $W$ and in the derived expressions for $F_\Phi$ and $F_X$, because $\exp(-aT)$ is suppressed relative to $W_{0,eff.}(\Phi)$ by a factor $1/(a\sigma_{min}(\varphi))$, which is typically ${\cal O}({1\over 10})$.

\subsubsection*{Avoiding decompactification during inflation}

We will now evaluate the constraint eq.~(\ref{fconstraint}), which must be satisfied to avoid decompactification. For this purpose, let us first focus on the region where $\Phi$ is large. Note that we want to produce a hierarchy in $W$, so we want $W_0$ to be many orders of magnitude smaller than the polynomial field terms (at least until the end of inflation, when the polynomial terms disappear). Combined with the fact $|X| \ll 1$, this allows us to approximate $W \approx \alpha b\Phi^n$. Furthermore, since $F_X$ has the highest power of $\Phi$ among the $F$-terms ($F_X \sim e^{K/2}\alpha\Phi^n$) $F_X$ dominates among the $F$-terms for large $\Phi$. Hence, eq.~(\ref{fconstraint}) becomes,
\begin{equation}
	\frac{\sqrt{|F_\Phi^2| + |F_X^2|}}{\sqrt{3}e^{K/2}|\langle W\rangle |} \sim \frac{|F_X|}{\sqrt{3}e^{K/2}|\langle W \rangle|} \sim \frac{e^{K/2}\alpha\Phi^n}{\sqrt{3}e^{K/2}\alpha b\Phi^n} \sim \frac{1}{\sqrt{3}b}.
\end{equation}
Therefore, at large $\Phi$, the ratio between the $F$-terms and $ \langle W \rangle$ \emph{is constant}. Thus, as long as we pick an ${\cal O}(1)$ value for $b$ such that the ratio is $\mathcal O(1)$, there is no danger of decompactification for large $\Phi$.

\begin{figure}[t!]
\centering
{\includegraphics[scale=0.8]{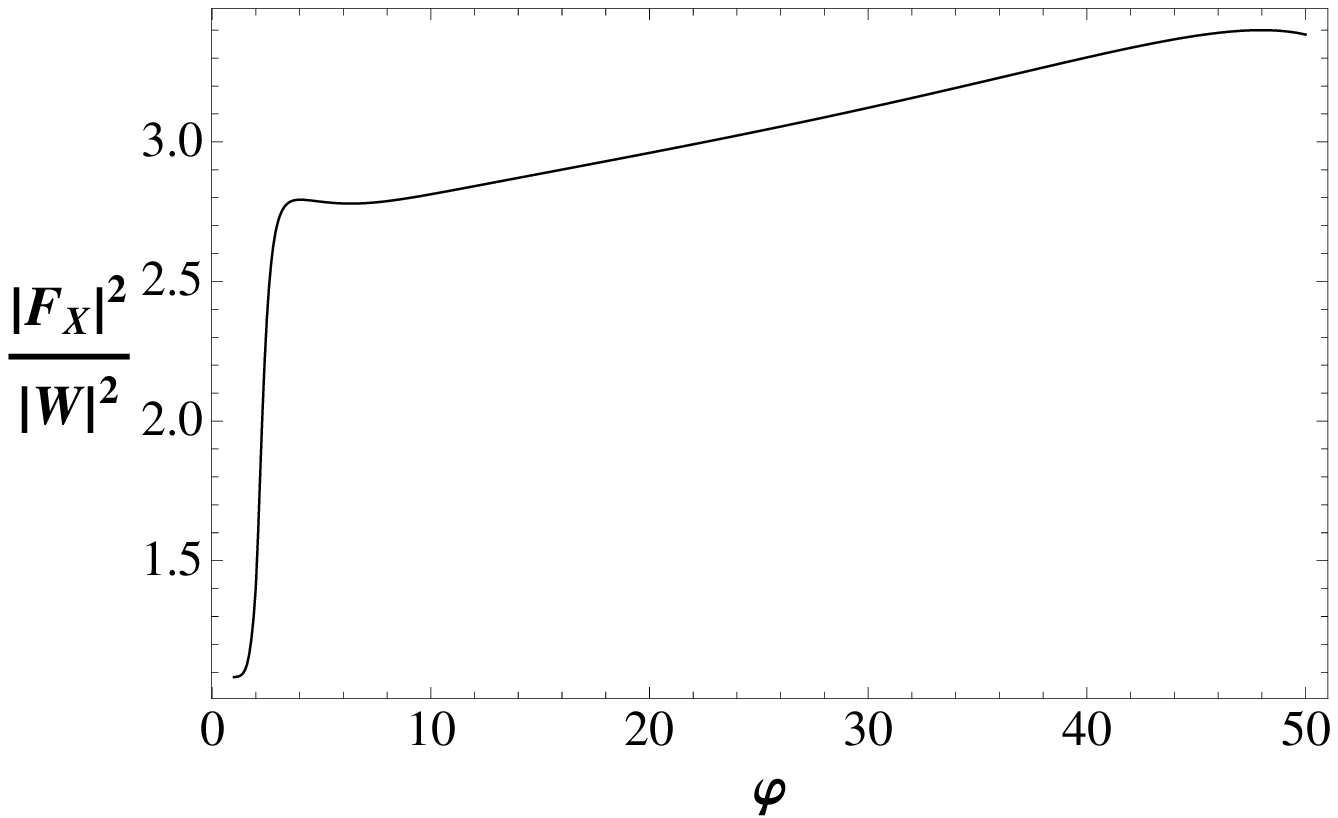}}\vspace*{2ex}
{\includegraphics[scale=0.8]{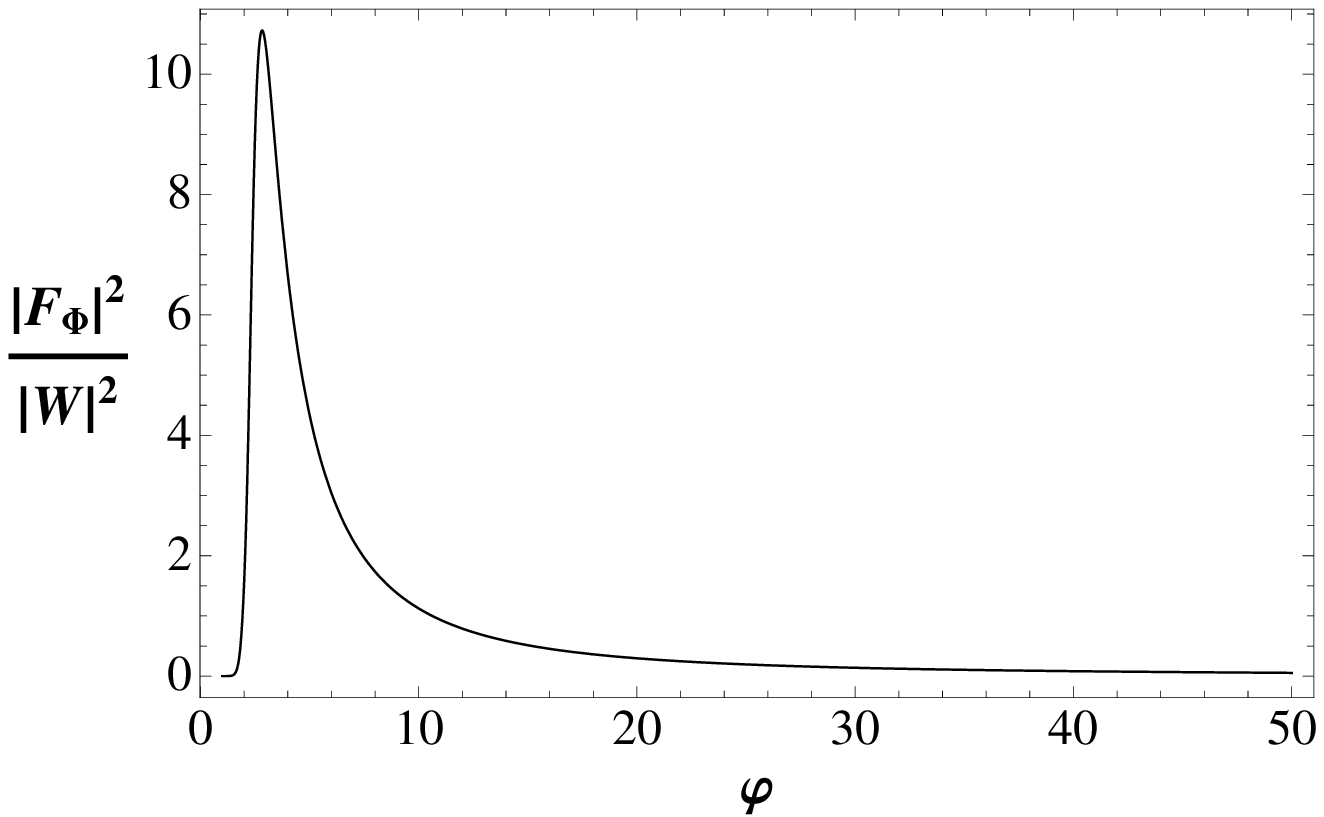}}
\caption{The $|F/W|^2$-ratios plotted as functions of the inflaton $\varphi$ with $X$ and $T$ adiabatically tracking their instantaneous minima.}
\label{fig.1}
\end{figure}

\begin{figure}[t!]
\centering
{\includegraphics[scale=0.8]{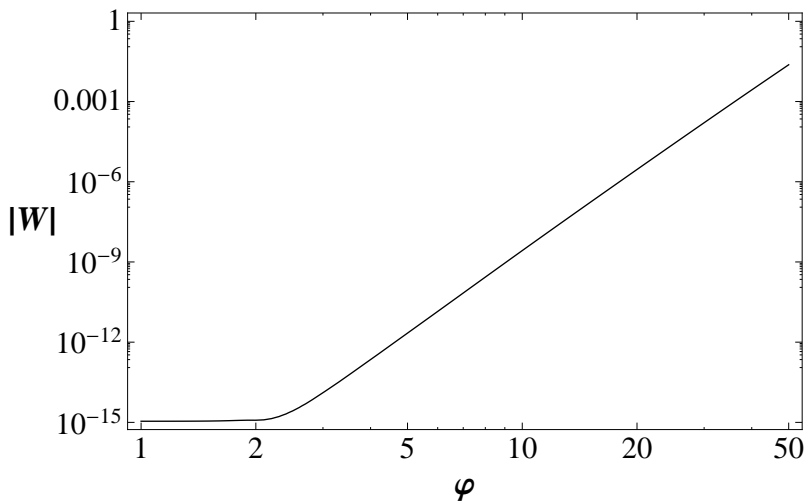}}
\caption{The vev of the superpotential $|\langle W\rangle|=|\langle W(\varphi,X(\varphi),T(\varphi))\rangle|$ plotted as a function of the inflaton $\varphi$ with $X$ and $T$ adiabatically tracking their instantaneous minima.}
\label{fig.2}
\end{figure}

We depict this behavior in Fig.~\ref{fig.1} for an exemplary choice of parameters in eq.~(\ref{KWdef}) given by: $A=1$, $a=\frac{2\pi}{10}$, $W_0=-10^{-15}$, $\alpha=5\times 10^{-19}$, $b=\sqrt{2/5}$, $n=10$, and $\gamma=2$.
This choice of parameters gives an effective inflationary potential $V(\varphi)\sim \varphi^{20}$ for $\varphi\lesssim \varphi_{60}\simeq 50 M_{\rm P}$ with the choice of $\alpha$ giving us $\delta\rho/\rho\simeq 1.6\times 10^{-5}$ at $\varphi_{60}$. Here we have approximated the functions $f,g$ in eq.~(\ref{KWdef}) by $f(X)=b+X+X^2/2$ and $g(X)=1+X$ for definiteness, to check explicitly that the higher-order terms do not spoil the behaviour of the model, as expected from the smallness of $X$ during inflation. Fig.~\ref{fig.2} shows us $|\langle W(\varphi,X(\varphi),T(\varphi))\rangle|$ as a function of the inflation $\varphi$, where $X(\varphi)$, $T(\varphi)$ denote the fields $X, T$ adiabatically tracking their instantaneous minima at every given value of $\varphi$.

Now, let us examine the region when $\Phi$ is small, i.e. sub-Planckian. In this region, the $F_\Phi$ term dominates as it has one smaller power of $\Phi$ than $F_X$. This means in the small $\Phi$ limit eq.~(\ref{fconstraint}) can be written as
\begin{equation}
\label{smallFconstraint}
	\frac{|F_\Phi|}{\sqrt{3}e^{K/2}|\langle W \rangle|} \sim \mathcal O(1).
\end{equation} 
Because $\eta$ vanishes, we can ignore the K\"ahler covariantization of the derivative in evaluating $F_\Phi$.  Hence, we are allowed to apply the global SUSY approximation, so 
\begin{equation}
\label{stuff}
  F_\Phi \approx e^{K/2}(n\alpha b\Phi^{n-1} + n\alpha X\Phi^{n-1}) \approx e^{K/2}n\alpha b\Phi^{n-1}~.
\end{equation}
We also dropped the $X\Phi^{n-1}$ term in (\ref{stuff}) since $\langle X\rangle\simeq 0$ (due to the $-\gamma (X\bar X)^2$ term in the K\"ahler potential). It follows that
\begin{equation}
\label{fwratio}
	\frac{|F_\Phi|}{\sqrt{3}e^{K/2}|\langle W \rangle|} \approx \frac{n\alpha b\Phi^{n-1}}{\sqrt{3}(\alpha b\Phi^n + W_0)}\sim\frac{1}{\Phi}\quad.
\end{equation}
We have again dropped the exponential in $T$ from $W$ since after inflation ends (and hence also close to the end of inflation), the non-perturbative term in the superpotential $\exp(-a T)$ at the minimum for $T$ is again smaller than $W_{0,eff.}(\Phi)$ by a factor $\sim 1/(aT)\ll1$ (and so can be neglected when evaluating the ratio eq.~(\ref{fwratio})). Furthermore, we again dropped the $X\Phi^n$ term for the same reason as above ($\langle X\rangle\simeq 0$ during inflation).

The $\Phi^{-1}$ scaling of the ratio eq.~(\ref{fwratio}) presents us with a danger of losing the $T$-minimum by producing too much uplifting in $|F_\Phi|^2$ \emph{after} the exit from inflation, for very small $\varphi\ll\varphi_{exit}$. To prevent this from happening, $F_{\Phi}$ in eq.~(\ref{fwratio}) needs to satisfy eq.~(\ref{smallFconstraint}) for all $\varphi$. This is easy to check;  the function $x^{n-1}/(x^n+c)$ has one global maximum for $x>c>0$. Thus,  we need
\begin{equation}
	\max\left(\frac{|F_\Phi|}{\sqrt{3}e^{K/2}|\langle W \rangle|}\right) \sim \mathcal O(1).
\end{equation}
Calculating the maximum of eq.~(\ref{fwratio}), our constraint becomes
\begin{equation}
\label{constraint}
	\max\left(\frac{|F_\Phi|}{\sqrt{3}e^{K/2}|\langle W \rangle|}\right) = \frac{n-1}{\sqrt{3}}\left(\frac{\alpha b}{W_0(n-1)}\right)^{1/n} \sim \mathcal O(1).
\end{equation}
If we take $b \sim \mathcal O(1)$, then we need $\alpha \sim W_0$. This can be satisfied in principle, even with the correct size of inflaton-generated density perturbations 
at the 60th e-folding,  by choosing $n>1$ sufficiently large.  We will now make this statement precise by discussing the hierarchies and perturbations that result for various values of $n$ and $W_0$.

\subsection{The structure of the $F$-terms and a no-go result}\label{nogo}

By inspecting the properties of the ratio eq.~(\ref{fwratio}) we are led towards two related observations.

Firstly, note that eq.~(\ref{fwratio}) is universal for all inflation models in which $F_\Phi$ is the dominant contribution to $V$ at small values of $\Phi$, i.e.
when the inflaton's own $F$-term dominates at small $\Phi$. This statement holds regardless of whether the model is small-field or large-field, since the analysis uses eq.~(\ref{fwratio}) only at small $\Phi$ close to the post-inflationary minimum, where Taylor expansion always gives us a polynomial form of the superpotential in $\Phi$. Thus the constraint on $n$ arising from these considerations is generic.\footnote{Non-generic exceptions include the case where the functional form of $W(\Phi)$ in the
small $\Phi$ regime is a single exponential, which by itself has no minimum and thus invalidates the preceding argument.  In any such scenario one needs to build in
a mechanism for graceful exit.}

We emphasize again that the constraint from eq.~(\ref{fwratio}) arises for very small inflaton values $|W_0|<\varphi\ll\varphi_{exit}$ \emph{after} inflation ended at $\varphi_{exit}\gg|W_0|$. Due to the desired smallness of $|W_0|$ in models with low-energy supersymmetry, this constraint therefore applies to both large-field and small-field models. 

Secondly, we observe that as long as the $F$-term driving inflation is given by $F_\Phi$, i.e. by the inflaton's own $F$-term, we see that eq.~(\ref{fwratio}) tells us that $F_\Phi$ decays relative to $W$ as $\Phi^{-1}$ for large $\Phi$. Due to the $-3e^K |W|^2$ term in the supergravity $F$-term scalar potential, this implies that $V(\Phi)$ is curving \emph{downwards} with \emph{increasing} $\partial^2V/\partial\Phi^2$ for very large values of $\Phi$, rendering a monotonically increasing inflaton potential at large field
values impossible. 

This second property of eq.~(\ref{fwratio}) leads us to add a second chiral field $X$ such that $\langle X\rangle\simeq 0$ during inflation but $|F_X|\gg |F_\Phi|$ with $|F_X/e^{K/2}W|\simeq const.\gtrsim \sqrt 3$ so that $V(\Phi)\sim |F_X|^2$ is monotonically increasing with $\Phi$ at large values of $\Phi$ (as one would need for large-field inflation).  One way to satisfy these requirements is to have a linear function of $X$ multiplying the inflationary polynomial in $\Phi$ inside $W$, which led to the choice of eq.~(\ref{KWdef}). 

Note that these latter considerations are similar to the ones which led to the first natural realization of $m^2\varphi^2$ chaotic inflation in 4D supergravity by Kawasaki, Yamaguchi \& Yanagida~\cite{sugrachaotic}. Our setup here shares the property with their model that inflation is driven by $F_X$ instead of $F_\Phi$, with $|F_X/e^{K/2}W|\simeq const.\gtrsim \sqrt{3}$ at large $\Phi$.  However, it generalizes the $X\Phi$ coupling in the superpotential used by these authors; this kind of coupling by itself would give  $\langle W\rangle\to 0$ at all $\Phi$ (due to the small $X$ vev), thus yielding a model
where the modulus $T$ would decompactify.

\subsection{Horse trading: Achievable hierarchies vs required inflaton power $n$ and $\delta\rho/\rho$}\label{whinnie}

In our model, $V$ is dominated at large $\varphi$ values by the $F_X$ term, as it has the largest power of $\varphi$. Hence, $F_X$ is the dominant term driving inflation, and we may approximate
\begin{equation}
	V \sim |F_X|^2 \sim \alpha^2\varphi^{2n}
\end{equation}
for large $\varphi$. Now, the magnitude of the density perturbation at 60 e-folds is given by
\begin{equation}
\label{delta}
	\delta \equiv \frac{\delta\rho}{\rho} = \left.\sqrt{\frac{1}{150\pi^2} \cdot \frac{V}{\epsilon}}\right|_{\varphi = \varphi_{60}},
\end{equation}
where $\epsilon = \frac{1}{2}\left(\frac{V'}{V}\right)^2$ is the $\epsilon$ slow roll parameter and $\varphi_{60}$ is the value of $\varphi$ at 60 e-folds. The measured value of $\delta$ is $2 \times 10^{-5}$. However, the purpose of this paper is not to create a fully realistic model compatible with experiment, but to demonstrate a mechanism
for circumventing the KL problem.
\begin{figure}[t!]
\centering
{\includegraphics[scale=1.05]{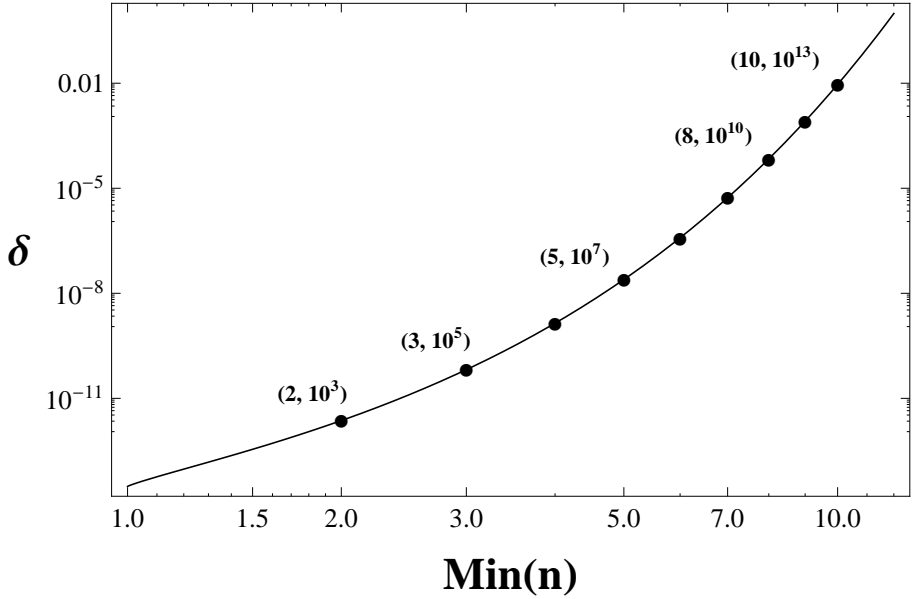}}
\caption{Minimum value of $n$ in the inflaton potential $V(\varphi)\sim \varphi^{2n}$ necessary to achieve a given $\delta\equiv \delta\rho/\rho$, at fixed $W_0=-10^{-15}$, satisfying the no-decompactification-constraint eq.~(\ref{fconstraint}). Points are labelled in the format $(n,\mathcal O(|W_i|/|W_0|))$, where $W_i=\langle W(\varphi_{60},X(\varphi_{60}),T(\varphi_{60}))\rangle$ is the initial superpotential (and ${\cal O}(|x|)$ here denotes the order of magnitude of $|x|$).}
\label{fig.3}
\end{figure}
For this reason, we leave $\delta$ as a free parameter and explore what values it can take in our model. It is easy to calculate $\varphi_{60}$ \cite{Baumann09} and in our case 
\begin{equation}
\label{60efolds}
	\varphi_{60} = 2\sqrt{60(n-1)}.
\end{equation}
Using the above three equations, we can solve for $\alpha$ as a function of $\delta$ and $n$, giving us
\begin{equation}
\label{alpha}
	\alpha = \frac{10\sqrt{3}\pi n\delta}{\varphi_{60}^{n+1}},
\end{equation}
where $\varphi_{60}$ is given by eq.~(\ref{60efolds}). Substituting this expression into eq.~(\ref{constraint}), we have
\begin{equation}
	(n-1)\left(\frac{10\sqrt{3}\pi bn\delta}{\varphi_{60}^{n+1}W_0(n-1)}\right)^{1/n} \leq 2\sqrt{3},
\end{equation} 
where we took $\mathcal O(1) = 2$ for calculational purposes. Now, in order for $m_{3/2} \sim \mathcal O(1) \unit{TeV}$, we take $W_0 = 10^{-15}$. Solving for $\delta$ as a function of $n$ yields
\begin{equation}
\label{delta-n}
	\delta \leq \frac{\left(\frac{2\sqrt{3}}{n-1}\right)^n\varphi_{60}^{n+1}W_0(n-1)}{10\sqrt{3}\pi bn}.
\end{equation}

This equation gives us the minimum value of $n$ needed to realize a given $\delta$ for a fixed value of $W_0$, subject to the constraint that decompactification does not occur. We have plotted $Min(n)$, the lower bound on $n$ necessary to achieve a given $\delta$ without too much uplifting, in Fig.~\ref{fig.3} for $b = \frac{1}{2}$. In each of the cases, the initial $W$ is many orders of magnitude larger than $W_0$, in the regime for large field inflation. After inflation, $W \approx W_0$, in the regime for TeV-scale SUSY breaking, thereby dynamically overcoming the KL problem. Furthermore, this toy model is robust in that it does not require extensive fine-tuning, since we may take $b$ to be any value in $\left[1/4\;,1/\sqrt{2}\,\right]$, and $\alpha$ is then determined by eq.~(\ref{alpha}). It is worth pointing out that to satisfy observational data on $\delta$, we require $n \geq 7$.

\begin{figure}[t!]
\centering
{\includegraphics[scale=1.05]{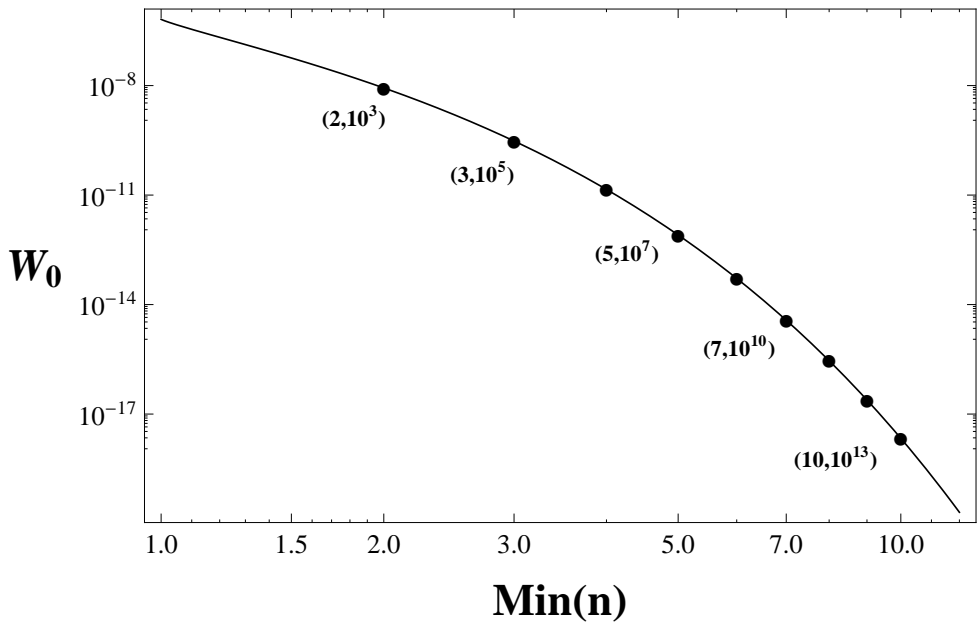}}
\caption{Minimum value of $n$ in the inflaton potential $V(\varphi)\sim \varphi^{2n}$ necessary to achieve a given post-inflationary vacuum VEV of the superpotential $W_0$, at fixed $\delta\equiv\delta\rho/\rho=2\times 10^{-5}$, satisfying the no-decompactification-constraint eq.~(\ref{fconstraint}). Points are labelled in the format $(n,\mathcal O(|W_i|/|W_0|))$, where $W_i=\langle W(\varphi_{60},X(\varphi_{60}),T(\varphi_{60}))\rangle$ is the initial superpotential (and ${\cal O}(|x|)$ here denotes the order of magnitude of $|x|$).}
\label{fig.4}
\end{figure}

We conclude this section by exploring the relation between $W_0$ and $n$ if we force $\delta  = 2 \times 10^{-5}$ (which is of interest as it is the observed value!).  We rearrange eq.~(\ref{alpha}) to get
\begin{equation}
  W_0 \geq \frac{10\sqrt{3}\pi nb \delta}{\left(\frac{2\sqrt{3}}{n-1}\right)^n\varphi_{60}^{n+1}(n-1)}.
\end{equation}
This equation gives us a lower limit for $W_0$ and hence gives us a lower limit for $m_{3/2}$ after inflation, and we have plotted the results in Fig.~\ref{fig.4}. 
Observational constraints on the spectral index and the tensor to scalar ratio (from e.g. \cite{WMAP}) require $2n \leq 4$.  Therefore,
$m_{3/2}$ cannot be on the order of the TeV scale in our toy model. On the other hand, even for small $n$, $\langle W \rangle$ decreases by several orders of magnitude during inflation, so $m_{3/2} \ll H$ is still satisfied in these models.

To summarize, while the class of $\varphi^{2n}$ chaotic inflation models we have studied in this section are able to accomodate very large ratios $H/m_{3/2}$ at large
enough $n$, this precise class of models is ruled out by experimental constraints on the scalar spectral index $n_s$ and the tensor to scalar ratio $r$ for all $n \geq 2$.    Therefore, 
the values of $n$ which are consistent with ${\rm TeV}$-scale supersymmetry breaking, are inconsistent with present cosmological data.  Some of the strongest
constraints on models in this subsection arise from the conditions required to prevent decompactification at small values of $\Phi$.
 For our large-field models, the small $\Phi$ regime has little to do with the period of inflation itself, and 
 it seems possible that by studying models with a slightly more
complicated exit from inflation (as in hybrid models \cite{hybrid}), one can build fully realistic theories with $m_{3/2} \sim {\rm TeV} \ll H$.

\section{Conclusions}

In this paper, we wrote down a class of toy models of inflation in string-inspired supergravity that successfully achieve $ m_{3/2} \ll  H$.  This shows that there is no general reason,
even in simple models of moduli stabilization in string theory, that the gravitino mass should be tied to the scale of inflation. Hence, the problem discussed
in \cite{Kallosh04} is not generic within inflation models in string-inspired supergravity constructions.  Rather, it is an artifact of studying very specific models.

However, our work leaves several important open questions.  Firstly, the models we presented are basically supergravity implementations of $\varphi^{2n}$ chaotic
inflation with various values of $n$.  The values of $n$ which we require to accomodate a ${\rm TeV}$ gravitino mass are ruled out by experiment (being clearly disfavored by
their predictions for both $n_s$ and $r$).   Values of $n$ which are consistent with cosmological data still yield a large hierarchy between $H$ and $m_{3/2}$, but it
is not large enough to allow ${\rm TeV}$ scale (or lower) gravitino mass.  Therefore, finding models which have $m_{3/2} \ll H$ and which are consistent with both precision cosmological data and
low-energy supersymmetry remains an open problem.
Secondly, our models are not derived in a top-down framework like string theory; the embedding of 
large-field inflation in string models with low-energy supersymmetry remains an
unmet challenge.  

It would be very interesting to try and address these problems, either in the class of models similar to \cite{KKLT}, or in alternatives based on e.g.
\cite{Conlon,eva,garyold,newone,newtwo,newthree,New,recentone,recenttwo}.    Our basic idea is to write down natural theories which allow $\langle W \rangle$ to vary by many orders of magnitude during
inflation, to free the Hubble scale of inflation from the final gravitino mass.  This idea should allow many different implementations, and it seems quite plausible that some of them will yield models consistent with both
cosmological observations and low-energy supersymmetry.

\bigskip
\centerline{\bf{Acknowledgements}}
\bigskip
We would like to thank R. Kallosh and A. Linde for bringing this problem to our attention, and G. Shiu, S. Trivedi and P. Yi for enjoyable informal discussions
in the autumn quarter of 2008.  We are indebted to L. Covi and L. McAllister for many insightful comments on a draft of this paper. This research was initiated at the  Aspen Center for Physics during the workshop ``Fingerprints of the
Early Universe" in the summer of 2009, and we are grateful to the Center and the workshop participants for providing a suitably stimulating atmosphere. 
S.K. is also grateful to the Mitchell Institute for Fundamental Physics and Astronomy at Texas A \& M for hospitality (and for hosting the stimulating Strings 2010
conference!) during the completion of this work. 
T.H. was supported in part by a Summer Research Fellowship (for undergraduate research) from the Stanford physics department.
The research of S.K. is supported in part by NSF grant PHY-05-51164.  The research of A.W. is supported by NSF grant PHY-0244728.

\end{document}